\documentclass[9pt,correspondence]{IEEEtran}
\usepackage{amsmath,amsfonts}
\usepackage{algorithm}
\usepackage{algpseudocode}

\usepackage[caption=false,font=normalsize,labelfont=sf,textfont=sf]{subfig}
\usepackage{textcomp}
\usepackage{stfloats}
\usepackage{url}
\usepackage{graphicx}
\usepackage{bm}
\def\BibTeX{{\rm B\kern-.05em{\sc i\kern-.025em b}\kern-.08em
		T\kern-.1667em\lower.7ex\hbox{E}\kern-.125emX}}
\usepackage{balance}
\usepackage{multirow}
\usepackage{times}

\def\plotsize{0.35\textwidth}

\begin{document}
\title{SALSA: A Sequential Alternating Least Squares Approximation Method For MIMO Channel Estimation}
\author{Sepideh Gherekhloo, Khaled Ardah, Martin Haardt  
\thanks{S. Gherekhloo and M. Haardt are with the Communications Research Laboratory (CRL), TU Ilmenau, Ilmenau, Germany (e-mail: \{sepideh.gherekhloo, martin.haardt\}@tu-ilmenau.de). K. Ardah is with Lenovo (Deutschland) GmbH (e-mail: kardah@lenovo.com).}  } 

%\markboth{Journal of \LaTeX\ Class Files,~Vol.~18, No.~9, September~2020}{}%

\maketitle

\begin{abstract}

In this paper, we consider the channel estimation problem in sub-6 GHz uplink wideband MIMO-OFDM communication systems, where a user equipment with a fully-digital beamforming structure is communicating with a base station having a hybrid analog-digital beamforming structure. A novel channel estimation method called Sequential Alternating Least Squares Approximation (SALSA) is proposed by exploiting a hidden tensor structure in the uplink measurement matrix. Specifically, by showing that any MIMO channel matrix can be approximately decomposed into a summation of $R$ factor matrices having a Kronecker structure, the uplink measurement matrix can be reshaped into a 3-way tensor admitting a Tucker decomposition. Exploiting the tensor structure, the MIMO channel matrix is estimated sequentially using an alternating least squares method. Detailed simulation results are provided showing the effectiveness of the proposed SALSA method as compared to the classical least squares method.

\end{abstract}


\section{Introduction}

\IEEEPARstart{M}assive MIMO \cite{marzetta2016} is one of the key enabling technologies of 5G-NR mobile communications \cite{Dahlman2018} and it shall remain relevant in future 6G wireless systems. By employing a large number of antennas at the base station (BS) relative to the number of scheduled users, massive MIMO systems increase the data throughput relative to legacy systems by providing a large beamforming gain and an improved multi-user interference suppression owing to its high spatial resolution \cite{EE_Emil}. Recently, massive MIMO communications have received a special attention with the introduction of millimeter-wave (mm-wave)-based wireless communications \cite{Heath2016}, since the use of massive MIMO in such systems becomes a requirement rather than an option to compensate the high pathloss encountered in the wireless communication systems at higher frequencies. However, it is well-known that the promised theoretical massive MIMO gains heavily rely on the availability of accurate channel state information (CSI) and the considered beamforming structure. 

On the one hand, classical fully-digital (FD) beamforming structures, which generally provide the maximum beamforming gain, require a dedicated radio frequency (RF) chain for each antenna element. This increases not only the implementation cost and complexity of massive MIMO systems, but also the circuit energy consumption. A promising solution to these issues relies on the recently introduced hybrid analog-digital (HAD) beamforming structures \cite{Ardah_UnifyDesign,Gao2016,Heath2016,Shokri2015,Gherekhloo_2020}, which use a combination of analog beamforming in the RF domain and digital beamforming in the baseband domain to reduce the number of RF chains as compared to FD beamforming structures, e.g., the number of RF chains can be as small as the number of transmitted data streams. 

On the other hand, in 5G-NR systems, for example, the BS estimates the CSI from uplink sounding reference signals (SRS) emitted by the user terminals (UEs). In mm-wave systems, the CSI estimation problem is often transformed into a multi-dimensional direction-of-arrival (DoA) estimation problem \cite{Alkhateeb2015,Ardah2020,ESPRITCE}, thanks to the low-rank (sparse) nature of mm-wave MIMO channels \cite{Heath2016}, where several techniques, e.g., compressed sensing \cite{Alkhateeb2015,Ardah2020} and ESPRIT \cite{ESPRITCE} can be readily employed to obtain a high CSI estimation accuracy while requiring a small number of training overhead. Differently, in sub-6 GHz-based systems, the MIMO channels often experience a high-rank nature, which makes most, if not all, mm-wave-based MIMO channel estimation methods unfeasible. To this end, classical channel estimation techniques, e.g., least-squares (LS) and minimum mean squared-error (MMSE) methods \cite{Barhumi2003,Biguesh2006} can be used to estimate sub-6 GHz-based MIMO channels. However, these methods were originally developed for single-antenna and small-scale MIMO systems and suffer from a severe performance degradation in difficult scenarios, e.g., with small number of training snapshots and/or a low signal-to-noise ratio (SNR). Since sub-6 GHz massive MIMO communications are, and will remain, an integral part of current and future wireless communication systems, more efficient channel estimation techniques than the classical methods are required. 

In this paper, we consider the channel estimation problem in sub-6 GHz uplink wideband MIMO-OFDM communication systems, where a single-user with a FD beamforming structure communicates with a BS having a HAD beamforming structure. By exploiting a hidden tensor structure in the uplink measurement matrix, we propose a novel channel estimation method called \textbf{S}equential \textbf{A}lternating \textbf{L}east \textbf{S}quares \textbf{A}pproximation (\textbf{SALSA}). Specifically, by showing that any MIMO channel matrix can be approximately decomposed into a summation of $R$ factor matrices having a Kronecker structure, the uplink measurement matrix can be reshaped into a 3-way tensor admitting a  Tucker decomposition \cite{Haardt2008}. Exploiting such a tensor representation, the MIMO channel matrix can be estimated sequentially using the classical ALS method \cite{Comon2009}. Detailed simulation results are provided showing that the SALSA-based approach can achieve a more accurate channel estimation in difficult scenarios as compared to the classical LS-based approach. 

\textbf{Notation:} The transpose, the complex conjugate, the conjugate transpose (Hermitian), and the Kronecker product are denoted as ${\bm A}^{\mathsf{T}}$, ${\bm A}^{*}$, ${\bm A}^{\mathsf{H}}$, and $\otimes$, respectively. Moreover, ${\bm I}_N$ is the $N\times N$ identity matrix, $\text{vec}\{ {\bm A} \}$ forms a vector by staking the columns of ${\bm A}$ over each other,  and the $n$-mode product of a tensor $\bm{\mathcal{A}}\in \mathbb{C}^{I_1\times I_2\times \dots,\times I_N}$ with a matrix ${\bm B}\in \mathbb{C}^{J\times I_n} $ is denoted as $\bm{\mathcal{A}} \times_n {\bm B}$.

\section{System Model}
\begin{figure}[t]
	\centering
	\includegraphics[width=0.95\linewidth]{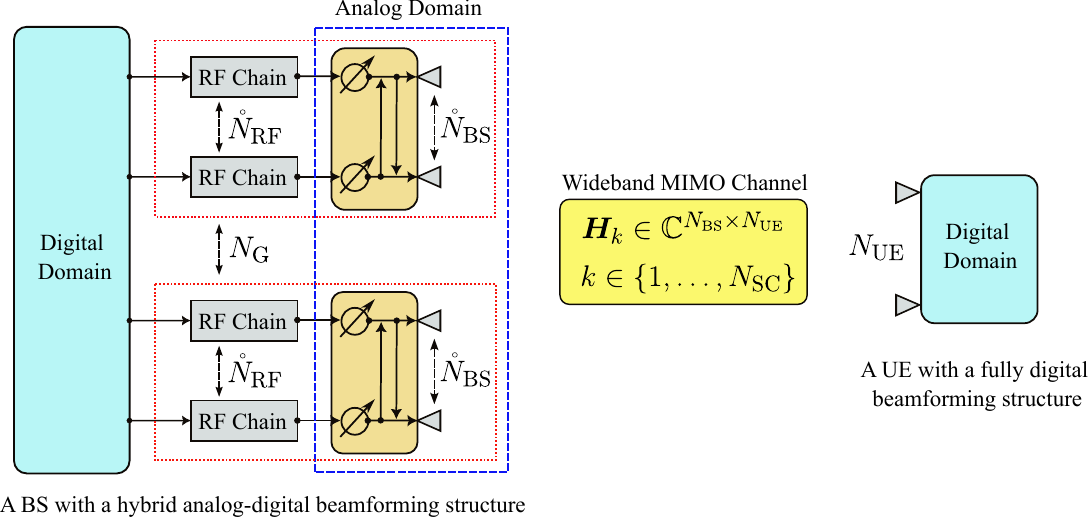}
	\caption{The considered uplink MIMO-OFDM communication system.}
	\label{fig:fig1}
\end{figure}

\begin{figure}[t]
	\centering
	\includegraphics[width=\plotsize]{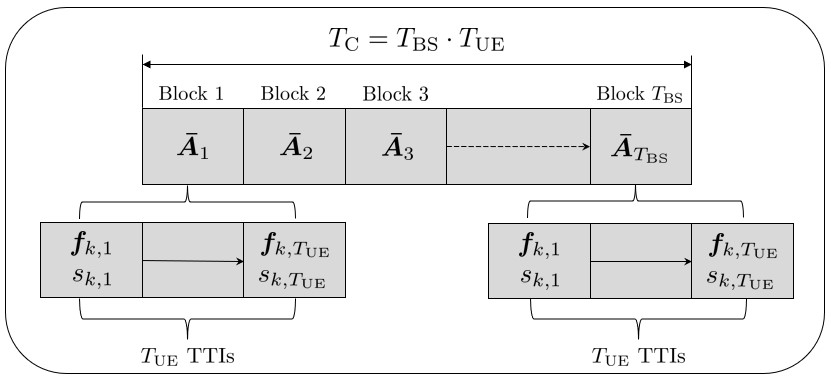}
	\caption{The channel coherence time $T_\text{C}$ division.}
	\label{fig:TTI}
\end{figure}

We consider an uplink single-user wideband MIMO-OFDM communication system, as depicted in Fig.~\ref{fig:fig1}, where a UE with $N_\text{UE}$ antennas is communicating with a BS with $N_\text{BS}$ antennas over $N_\text{SC}$ subcarriers. The UE has a FD beamforming structure while the BS has a HAD beamforming structure with $N_\text{RF} \leq N_\text{BS}$ radio-frequency (RF) chains. We assume that the $N_\text{BS}$ antennas and the $ N_\text{RF}$ RF chains are divided \textit{equally}\footnote{To simplify the exposition, we assume that $N_\text{BS}$, $N_\text{RF}$, and $N_\text{G}$ are selected so that $\mathring{N}_\text{BS}$ and $\mathring{N}_\text{RF}$ are integer numbers, without loss of generality.} into $ N_\text{G} \geq 1$ groups, where each group has $\mathring{N}_\text{BS} = \frac{N_\text{BS} }{N_\text{G}}$ antennas and $ \mathring{N}_\text{RF} = \frac{N_\text{RF}}{N_\text{G}}$ RF chains (i.e., $ N_\text{BS}  = N_\text{G}\cdot \mathring{N}_\text{BS}$ and $ N_\text{RF} = N_\text{G} \cdot  \mathring{N}_\text{RF}$) and the RF chains in every group are connected with every antenna element in the same group. Moreover, we assume a block-fading channel model as shown in Fig.~\ref{fig:TTI}, where the channel coherence-time $T_\text{C}$ is divided into $T_\text{BS} T_\text{UE}$ transmission time intervals (TTIs), i.e., every block has $T_{\text{UE}}$ snapshots. 

Let $\bm{\bar{A}}_i \in \mathbb{C}^{N_\text{BS} \times N_\text{RF}}$ denote the analog combining matrix at the $i$th block at the BS. Then, according to our above assumptions, $\bm{\bar{A}}_i$ has a block-diagonal structure given as\footnote{Note that if $N_\text{G} = 1$, the above analog structure coincides with the known {fully-connected} analog structure \cite{Ardah_UnifyDesign}, where every RF chain is connected to every antenna element. On the other hand, if $N_\text{G} = N_\text{RF}$, the above analog structure coincides with the known {partially-connected} analog structure \cite{Ardah_UnifyDesign}, where every RF chain is connected to a unique subset of antenna elements.} 
%\begin{align}\label{scfc}
%{\bm \bar A} = \frac{1}{\sqrt{\mathring{N}_\text{BS}}} \cdot \text{blockdiag}\{{\bm \tilde A}_1, \dots, {\bm \tilde A}_{G}\} \in \mathbb{C}^{N_\text{BS} \times N_\text{RF}},
%\end{align}
\begin{align}\label{scfc}
	\bm {\bar{A}}_i = \frac{1}{\sqrt{\mathring{N}_\text{BS}}} \cdot \begin{bmatrix}
		\bm{\bar{A}}_{i,1} & \dots & {\bm 0} \\
		\vdots & \ddots & \vdots \\
		{\bm 0} & \dots & \bm{\bar{A}}_{i,N_\text{G}}
	\end{bmatrix} \in \mathbb{C}^{N_\text{BS} \times N_\text{RF}},
\end{align}
where $\bm{\bar{A}}_{i,g} \in \mathbb{C}^{\mathring{N}_\text{BS} \times \mathring{N}_\text{RF}}$ is the $g$th block-matrix with  constant modulus entries, i.e., $\big|[\bm{\bar{A}}_{i,g}]_{[r,c]}\big| = 1$, where $[\bm{\bar{A}}_{i,g}]_{[r,c]}$ is the $(r,c)$th entry of $\bm{\bar{A}}_{i,g}$. 

The received signal by the BS in the $(i,j)$th TTI over the $k$th subcarrier, with $i \in \{1,\dots,T_\text{BS}\}$, $j \in \{1,\dots,T_\text{UE}\}$, $k \in \{1,\dots,N_\text{SC}\}$, can be expressed as
\begin{align}\label{Yun}
	\bm{\bar{y}}_{k,i,j} = \bm{\bar{A}}^{\mathsf{H}}_{i}  {\bm H}_k {\bm f}_{k,j} s_{k,j} + \bm{\bar{A}}^{\mathsf{H}}_{i} \bm{\bar{z}}_{k,i,j} \in \mathbb{C}^{N_\text{RF}},
\end{align}
where ${\bm f}_{k,j} \in \mathbb{C}^{{N}_\text{UE}}$ is the $(k,j)$th precoding vector, $s_{k,j} \in \mathbb{C}$ is the corresponding training symbol, $\bm{\bar{z}}_{i,j} \in \mathbb{C}^{{N}_\text{BS}}$ is the BS additive white Gaussian noise with zero mean and variance $\sigma_n^2$, and ${\bm H}_k \in \mathbb{C}^{{N}_\text{BS} \times {N}_\text{UE}}$ is the $k$th subcarrier frequency-domain MIMO channel matrix. 

Initially, we collect the measurement vectors $\{\bm{\bar{y}}_{k,i,j}\}_{j = 1}^{T_\text{UE}}$ next to each other as $\bm{\bar{Y}}_{k,i} = [\bm{\bar{y}}_{k,i,1}, \dots, \bm{\bar{y}}_{k,i,T_\text{UE}}]$, which can be written as  
 \begin{align}\label{Ybar0}
 \bm{\bar{Y}}_{k,i} = \bm{\bar{A}}^{\mathsf{H}}_{i}  {\bm H}_k {\bm F}_k + \bm{\bar{A}}^{\mathsf{H}}_{i} \bm{\bar{Z}}_{k,i} \in \mathbb{C}^{{N}_\text{RF} \times T_\text{UE}},
 \end{align}
where ${\bm F}_k = [{\bm f}_{k,1}s_{k,1}, \dots, {\bm f}_{k,T_\text{UE}} s_{k,T_\text{UE}} ] \in \mathbb{C}^{{N}_\text{UE} \times T_\text{UE}}$ and $\bm{\bar{Z}}_{k,i} = [\bm{\bar{z}}_{k,i,1}, \dots, \bm{\bar{z}}_{k,i,T_\text{UE}}]$. We assume that ${\bm F}_k,\forall k$, are designed with orthonormal rows, i.e., ${\bm F}_k {\bm F}^{\mathsf{H}}_k = {\bm I}_{N_\text{UE}}, \forall k$, and $T_\text{UE} \geq N_\text{UE}$. After applying the right-filtering to (\ref{Ybar0}) we obtain
\begin{align}\label{Yi}
	{\bm Y}_{k,i} = \bm{\bar{Y}}_{k,i} {\bm F}^{\mathsf{H}} = \bm{\bar{A}}^{\mathsf{H}}_{i}  {\bm H}_k + {\bm Z}_{k,i} \in \mathbb{C}^{N_\text{RF} \times N_\text{UE}},
\end{align}
where ${\bm Z}_{k,i} = \bm{\bar{A}}^{\mathsf{H}}_{i} \bm{\bar{Z}}_{k,i} {\bm F}^{\mathsf{H}}_k$. Next, we collect the measurement matrices $\{{\bm Y}_{k,i}\}_{i = 1}^{T_\text{BS}}$ on the top of each other as ${\bm Y}_k = \big[{\bm Y}^{\mathsf{T}}_{k,1}, \dots, {\bm Y}^{\mathsf{T}}_{k,T_\text{BS}}  \big]^{\mathsf{T}}$, which can be written as 
\begin{align}
	{\bm Y}_k = {\bm A}  {\bm H}_k + {\bm Z}_k \in \mathbb{C}^{L \times N_\text{UE}},
\end{align}
where $L = T_\text{BS} {N}_\text{RF}$, ${\bm A} = \big[\bm{\bar{A}}_{1}, \dots, \bm{\bar{A}}_{T_\text{BS}}  \big]^{\mathsf{H}} \in \mathbb{C}^{ L  \times N_\text{BS}}$, and ${\bm Z}_k = \big[{\bm Z}^{\mathsf{T}}_{k,1}, \dots, {\bm Z}^{\mathsf{T}}_{k,T_\text{BS}}  \big]^{\mathsf{T}}$. After that, we collect the measurement matrices $\{{\bm Y}_k \}_{k = 1}^{N_\text{SC}}$ next to each other as $\bm{{Y}} = [{\bm Y}_1, \dots, {\bm Y}_{N_\text{SC}}]$, which can be written as 
\begin{align}\label{Y}
	{\bm Y} = {\bm A}  {\bm H} + {\bm Z} \in \mathbb{C}^{L \times N_\text{UE} N_\text{SC}},
\end{align}
where ${\bm Z} = \big[{\bm Z}_{1}, \dots, {\bm Z}_{N_\text{SC}}  \big]$ and ${\bm H} = \big[\bm{H}_{1}, \dots, \bm{H}_{N_\text{SC}}  \big] \in \mathbb{C}^{ N_\text{BS} \times N_\text{UE} N_\text{SC}}$ is the total MIMO channel matrix.

\textbf{The baseline LS-based channel estimation method}: Given the measurement matrix in (\ref{Y}), a least-squares (LS)-based method can be used to obtain an estimate of the total MIMO channel matrix as
\begin{align}\label{LS}
	\bm{\hat{H}}_{\text{LS}} = [{\bm A}]^{+} {\bm Y} = \big[\bm{\hat{H}}_{1}, \dots, \bm{\hat{H}}_{N_\text{SC}}  \big] \in \mathbb{C}^{N_\text{BS} \times N_\text{UE} N_\text{SC}},
\end{align}
where $[\cdot]^+$ denotes the Moore-Penrose pseudo-inverse. 
%, i.e., $[{\bm A}]^{+} = {\bm A}^{\text{H}} ({\bm A} {\bm A}^{\text{H}})^{-1}$, if $T_\text{BS} {N}_\text{RF} <  N_\text{BS}$ and $[{\bm A}]^{+} = ({\bm A}^{\text{H}} {\bm A})^{-1} {\bm A}^{\text{H}}$ , if $T_\text{BS} {N}_\text{RF}  \geq  N_\text{BS}$. 
Note that, due to the left filtering, the LS-based method requires that $L  \geq  N_\text{BS}$, i.e.,  $T_\text{BS} \geq \frac{N_\text{BS}} { N_\text{RF}}$ to provide an accurate channel estimate.

\section{The proposed SALSA method}

To obtain a more accurate channel estimate while reducing the training overhead, we propose in this section a novel channel estimation method called \textbf{SALSA}, which is derived by exploiting a hidden tensor structure in the measurement matrix in (\ref{Y}). To show this, we first recall the following propositions from \cite{VanLoan1993,King2016,Garvey2018}.

\textbf{Proposition 1:} Let ${\bm X}$ be a matrix given as 
\begin{align}
	{\bm X} = {\bm X}_1 \otimes {\bm X}_2 =  \begin{bmatrix}
		{\bm X}_{1,1} & \dots & {\bm X}_{1,J_1} \\ 
		\dots \\
		{\bm X}_{I_1,1} & \dots & {\bm X}_{I_1,J_1} \\ 
	\end{bmatrix} \in \mathbb{C}^{I \times J},
\end{align}
where ${\bm X}_1  \in \mathbb{C}^{I_1 \times J_1} $, ${\bm X}_2  \in \mathbb{C}^{I_2 \times J_2}$,  $I = I_1  I_2$, $J = J_1  J_2$, and ${\bm X}_{n,m} = [{\bm X}_1]_{[n,m]} {\bm X}_2$ is the $(n,m)$th block-matrix of ${\bm X}$. Let ${\bm K} \in \mathbb{C}^{I_1J_1 \times I_2J_2}$ be a rank-one matrix given as  
\begin{align}
	{\bm K}  = \begin{bmatrix}
		\text{vec}\{{\bm X}_{1,1}\}^{\text{T}}  \\ 
		\dots \\
		\text{vec}\{{\bm X}_{I_1,1}\}^{\text{T}}  \\ 
		\dots \\
		\text{vec}\{{\bm X}_{1,J_1}\}^{\text{T}}  \\ 
		\dots \\
		\text{vec}\{{\bm X}_{I_1,J_1}\}^{\text{T}}  \\ 
	\end{bmatrix} = \text{vec}\{ {\bm X}_1  \}  \text{vec}\{ {\bm X}_2  \}^{\text{T}},
\end{align}
with the rank-one truncated-SVD given as  ${\bm K} = \sigma {\bm u} {\bm v}^{\text{H}}$,  where ${\bm u} \in \mathbb{C}^{I_1J_1}$ and ${\bm v} \in \mathbb{C}^{I_2J_2}$ are the left and right singular vectors of ${\bm K}$, respectively, and $\sigma$ is the associated singular value. Then, the optimal solution to 
\begin{align}
	\underset{ {\bm X}_1 , {\bm X}_2}{\text{minimize }} \Vert {\bm X} - \big({\bm X}_1 \otimes {\bm X}_2 \big) \Vert^2_\text{F} 
\end{align}
can be obtained as
\begin{align}
	{\bm X}_1 &= \text{reshape}\{ \sqrt{\sigma} {\bm u}, I_1, J_1 \} \label{X1} \\
	{\bm X}_2 &= \text{reshape}\{ \sqrt{\sigma} {\bm v}^{*}, I_2, J_2 \} \label{X2}.
\end{align} 
\textbf{Proof}: Please refer to \cite{King2016} for more details.  
 
\textbf{Proposition 2:} For any given $I \times J$ matrix ${\bm X}$, it can be \textit{approximately} written as a summation of $R\geq1$ factor matrices as
\begin{align}
	{\bm X}= \sum_{r = 1}^{R} {\bm X}_r = \sum_{r = 1}^{R} {\bm X}_{1,r} \otimes {\bm X}_{2,r},
\end{align}
where ${\bm X}_r = {\bm X}_{1,r} \otimes {\bm X}_{2,r}$, ${\bm X}_{1,r} \in \mathbb{C}^{I_1 \times J_1} $, and ${\bm X}_{2,r}  \in \mathbb{C}^{I_2 \times J_2} $, $I = I_1   I_2$, and $J = J_1   J_2$. 

\textbf{Proof}: The proof follows directly by applying {Proposition~1} sequentially \cite{Garvey2018}. The corresponding Proposition is summarized in Algorithm \ref{KronAprx}.

\begin{figure}
	\centering
	\includegraphics[width=0.7\linewidth]{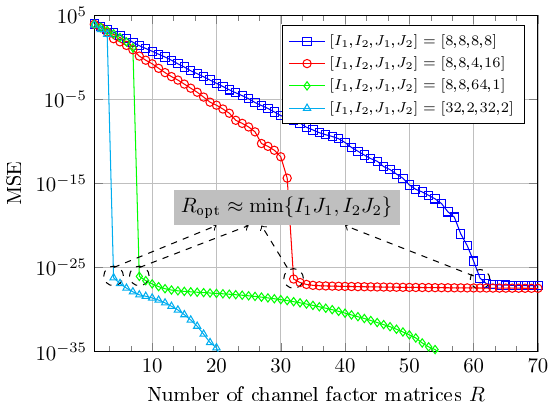}
	\caption{MSE vs. the number of channel factor matrices $R$ assuming $N_\text{BS} = 64$, $N_\text{UE} = 4$, and $N_\text{SC} = 16$, where $\text{MSE} = \Vert {\bm H} - \sum_{r = 1}^{R} {\bm C}_{r} \otimes {\bm B}_{r} \Vert^2_\text{F}$. Here, the total MIMO channel matrix ${\bm H} = \big[\bm{H}_{1}, \dots, \bm{H}_{N_\text{SC}}  \big] \in \mathbb{C}^{ I \times J}$ is generated following the 3GPP CDL channel model \cite{3gpp38901,CDLChannel} with the main system parameters outlined in Table~\ref{TableSys}. Please refer to Section \ref{SecNumResults} for more details.}
	\label{fig2}
	\vspace{-10pt}
\end{figure}

Let $I = {N}_\text{BS}$ and $J = {N}_\text{UE}N_\text{SC}$. Then, from Proposition 2, the total frequency-domain MIMO channel matrix ${\bm H} \in \mathbb{C}^{ I \times J}$ in (\ref{Y}) can be \textit{approximately} written as 
\begin{align}\label{HCB}
{\bm H} \approx \sum_{r = 1}^{R} {\bm C}_{r} \otimes {\bm B}_{r} \in \mathbb{C}^{I \times J},
\end{align}
where ${\bm B}_r \in \mathbb{C}^{I_1 \times J_1}$, ${\bm C}_r \in \mathbb{C}^{I_2 \times J_2}$, $I = I_1I_2$, and $J = J_1J_2$. As shown in Fig. \ref{fig2}, the \textit{approximation} becomes tighter as the number of channel factor matrices $R$ increases. More importantly, we can see that in case of full rank channels, the optimal value of $R$, denoted in the figure by $R_\text{opt}$, is dependent on the division scenario of $I$ and $J$, where $R_\text{opt} \approx \min\{I_1J_1, I_2J_2\}$. In other words, reducing the dimension of one of the channel factor matrices, i.e., ${\bm B}_r \in \mathbb{C}^{I_1 \times J_1}$ or ${\bm C}_r \in \mathbb{C}^{I_2 \times J_2}$, reduces the value of $R_\text{opt}$.      

\begin{algorithm}[t] 
	\caption{Sequential Kronecker Factorization}
	\label{KronAprx}
	\begin{algorithmic}[1]
		\State{\textbf{Input}: A matrix ${\bm X} \in \mathbb{C}^{I \times J}$ }
		\State{Select $R$, $I_1 $, $ J_1$, $I_2 $, $ J_2$ such that $I = I_1  I_2$ and $J = J_1  J_2$}
		
		\For{$r = 1$ to $R$}
		\State{Get ${\bm X}_{r} = {\bm X} -  \sum_{r' = 1}^{r-1} {\bm X}_{1,r'} \otimes {\bm X}_{2,r'} $}
		
		\State{Given ${\bm X}_{r}$, get ${\bm X}_{1,r}$ and ${\bm X}_{2,r}$ using (\ref{X1}) and (\ref{X2}), respectively}	
		
		\EndFor
		\State{\textbf{Output}: ${\bm \hat X} = \sum_{r = 1}^{R} {\bm X}_{1,r} \otimes {\bm X}_{2,r} \in\mathbb{C}^{I  \times  J }$ }
	\end{algorithmic}
\end{algorithm}

Let $L = T_\text{BS} {N}_\text{RF}$. Then, by substituting (\ref{HCB}) into (\ref{Y}), and assuming $R$ is sufficiently large, we can write 
\begin{align}\label{Ybar}
	{\bm Y} &= {\bm A} \Bigg( \sum_{r = 1}^{R} {\bm C}_{r} \otimes {\bm B}_{r} \Bigg) +  {\bm Z}  
	 =\sum_{r = 1}^{R} {\bm A} ({\bm C}_{r} \otimes {\bm B}_{r}) + {\bm  Z} \nonumber \\
	&= \sum_{r = 1}^{R} 	{\bm  Y}_{r}  + {\bm  Z} \in\mathbb{C}^{ L   \times J }, 
\end{align}
where ${\bm  Y}_{r} = {\bm A} ({\bm C}_{r} \otimes {\bm B}_{r}) \in\mathbb{C}^{ L  \times J }$. 
From (\ref{Ybar}), we note that ${\bm Y}_{r} $ can be seen as the 1-mode unfolding of a 3-way Tucker tensor given as \cite{Haardt2008} 
\begin{align}\label{Yrten}
	\bm{ \mathcal{Y}}_{r} = \bm{\mathcal{S}} \times_1  {\bm A}  \times_2  {\bm B}^{\mathsf{T}}_{r}  \times_3  {\bm C}^{\mathsf{T}}_{r}   \in\mathbb{C}^{ L  \times J_1 \times J_2 }, 
\end{align}
where $\bm{\mathcal{S}} \in\mathbb{Z}^{ I \times I_1 \times  I_2  } $ is the core-tensor with the 1-mode unfolding given as $[\bm{\mathcal{S}}]_{(1)} \overset{\text{def}}{=} {\bm I}_{I}$. The $\ell$-mode unfolding of $\bm{ \mathcal{Y}}_{r}$, $\ell = \{1,2,3\}$, can be expressed as 
\begin{align}
	[\bm{ \mathcal{Y}}_{r}]_{(1)} &= {\bm A} [\bm{\mathcal{S}}]_{(1)} ({\bm C}_{r} \otimes {\bm B}_{r}) \in\mathbb{C}^{ L  \times J  }, \\
	[ \bm{ \mathcal{Y}}_{r}]_{(2)} &= {\bm B}^{\mathsf{T}}_{r}  [\bm{\mathcal{S}}]_{(2)} ({\bm C}_{r} \otimes {\bm A}^{\mathsf{T}}) \in\mathbb{C}^{ J_1 \times L  J_2 }, \\
	[\bm{ \mathcal{Y}}_{r}]_{(3)} &= {\bm C}^{\mathsf{T}}_{r}  [\bm{\mathcal{S}}]_{(3)} ({\bm B}_{r} \otimes {\bm A}^{\mathsf{T}}) \in\mathbb{C}^{ J_2 \times L  J_1  }.
\end{align}

From (\ref{Yrten}), the 3-way Tucker tensor form of (\ref{Ybar}) can be expressed as  
\begin{align}\label{Yten}
	\bm{ \mathcal{Y}} = \sum_{r = 1}^{R} \bm{ \mathcal{Y}}_{r}  + \bm{ \mathcal{Z}}  \in\mathbb{C}^{ L   \times J_1 \times J_2 },
\end{align}
where $\bm{{{ \mathcal{Z} }}}$ is the 3-way tensor representation of the noise matrix ${\bm  Z} $. This latter formulation suggests that the factor matrices $\{{\bm B}_{r}, {\bm C}_{r}\}_{r = 1}^{R}$ can be estimated sequentially as follows. Let $\bm{ \mathcal{Y}}_{r}$ be the tensor obtained at the $r$th sequential step as
\begin{align}
	\bm{ \mathcal{Y}}_{r} = \bm{ \mathcal{Y}} - \sum_{r' = 1}^{r-1} \bm{\mathcal{Y}}_{r'} \in\mathbb{C}^{ L  \times J_1 \times J_2 }.
\end{align} 
%where $\bm{\bar \mathcal{Y}}_{u,i}$ are the tensors obtained from the factor matrices $\{{\bm B}_{u,i}, {\bm C}_{u,i}\}_{i = 1}^{r-1}$. 

Then, by exploiting the 2-mode and the 3-mode unfoldings, the $r$th factor matrices ${\bm B}_{r} $ and $ {\bm C}_{r}$ can be obtained using, e.g., the ALS method \cite{Comon2009}, where one factor matrix is assumed to be fixed when solving for the other. Specifically, ${\bm B}_{r} $ and $ {\bm C}_{r} $ can be obtained as   
\begin{align}
{\bm B}^{\mathsf{T}}_{r} & = [\bm{  \mathcal{Y}}_{r}]_{(2)} \big[ \bm{\Psi}_{2} \big]^{+} 
 = [\bm{  \mathcal{Y}}_{r}]_{(2)}  {\bm \Psi}_{2}^{\text{H}} [{\bm \Psi}_{2}{\bm \Psi}_{2}^{\text{H}} ]^{-1}  \label{bc}\\
{\bm C}^{\mathsf{T}}_{r} &=  [\bm{ \mathcal{Y}}_{r}]_{(3)} \big[ {\bm \Psi}_{3}\big]^{+}   = [\bm{  \mathcal{Y}}_{r}]_{(3)}  {\bm \Psi}_{3}^{\text{H}} [{\bm \Psi}_{3} {\bm \Psi}_{3}^{\text{H}} ]^{-1} \label{cc},
\end{align}
where ${\bm \Psi}_{2}$ and ${\bm \Psi}_{3}$ are given as 
\begin{align}
{\bm \Psi}_{2}& =  [\bm{\mathcal{S}}]_{(2)} ({\bm C}_{r} \otimes {\bm A}^{\mathsf{T}}) \in \mathbb{C}^{ I_1 \times L  J_2  } \\
{\bm \Psi}_{3} &= [\bm{\mathcal{S}}]_{(3)} ({\bm B}_{r} \otimes {\bm A}^{\mathsf{T}}) \in \mathbb{C}^{ I_2 \times L  J_1  }.
\end{align}

%On the other hand, to reduce the noise impact, the WALS method \cite{Giampouras2019} can be used instead to obtain ${\bm B}_{r} $ and $ {\bm C}_{r}$ as
%\begingroup\makeatletter\def\f@size{10}\check@mathfonts
%\def\maketag@@@#1{\hbox{\m@th\small\normalfont#1}}% 
%\begin{align} 
%	{\bm B}^{\mathsf{T}}_{r} &=  [\bm{  \mathcal{Y}}_{r}]_{(2)}   {\bm \Psi}_{2}^{\mathsf{H}}  \big[ {\bm \Psi}_{2}  {\bm \Psi}_{2}^{\mathsf{H}}  + \lambda {\bm D}_2     \big]^{-1}  \label{bw} \\ 
%	{\bm C}^{\mathsf{T}}_{r} &=   [\bm{  \mathcal{Y}}_{r}]_{(3)}   {\bm \Psi}^{\mathsf{H}}_{3}  \big[ {\bm \Psi}_{3} {\bm \Psi}^{\mathsf{H}}_{3}  + \lambda {\bm D}_3 \big]^{-1} \label{cw}, 
%\end{align}\endgroup    
%where $ {\bm D}_2 $ and ${\bm D}_3$ are diagonal matrices given as 
%\begingroup\makeatletter\def\f@size{9}\check@mathfonts
%\def\maketag@@@#1{\hbox{\m@th\small\normalfont#1}}% 
%\begin{align}
%	& {\bm D}_2 =\text{diag}\{ \Vert  [{\bm \Psi}_{2}]_{[1,:]} \Vert, \dots,  \Vert  [{\bm \Psi}_{2}]_{[{N^{\text{B}}_{\text{BS}}},:]} \Vert  \} \in \mathbb{R}^{N^{\text{B}}_{\text{BS}} \times N^{\text{B}}_{\text{BS}}}   \label{D1} \\
%	& {\bm D}_3 = \text{diag}\{ \Vert  [{\bm \Psi}_{3}]_{[1,:]} \Vert, \dots,  \Vert  [{\bm \Psi}_{3}]_{[N^{\text{C}}_{\text{BS}},:]} \Vert  \} \in \mathbb{R}^{N^{\text{C}}_{\text{BS}} \times N^{\text{C}}_{\text{BS}}} , 	\label{D2} 
%\end{align}\endgroup
%where $[{\bm \Psi}_{x}]_{[k,:]}$ is the $k$th row vector of ${\bm \Psi}_{x}$ and $\lambda$ is the weighting parameter \textcolor{blue}{which is a} function of the noise variance $\sigma_n^2$. 

Algorithm~\ref{SALSA} summarizes the proposed SALSA method for estimating the total MIMO channel matrix ${\bm H} \in\mathbb{C}^{I \times J}$, which is guaranteed to converge monotonically to, at least, a local optimum solution \cite{Comon2009}.    
 
\begin{algorithm}
	\caption{SALSA For MIMO-OFDM Channel Estimation}
	\label{SALSA}
	\begin{algorithmic}[1]
		\State{\textbf{Input}: Measurement matrix ${\bm Y} \in \mathbb{C}^{ L \times J}$ as in (\ref{Y})}
		\State{Select $R \geq 1$, $N_{\text{max-iter}} \geq 1$, $I_1, I_2, J_1 $, and $ J_2$ such that $I = I_1I_2 = {N}_\text{BS}$ and  $J = J_1J_2 = {N}_\text{UE}N_\text{SC}$}
		\State{Obtain the 3-way Tucker tensor $\bm{ \mathcal{Y}}$ in (\ref{Yten}) from ${\bm Y}$}
			%, i.e., $$\text{\textit{In MATLAB:} }\bm{ \mathcal{Y}} = \text{reshape}\{ {\bm Y} ,L, J_1, J_2 \}$$} 
		\For{$r = 1$ to $R$}
		\State{Get $\bm{ \mathcal{Y}}_{r} = \bm{\mathcal{Y}} -  \sum_{r' = 1}^{r-1}\bm{\mathcal{ \hat Y}}_{r'} $}\label{Step3}

		\State{Initialize ${\bm C}^{(0)}_{r} \in \mathbb{C}^{I_2 \times J_2}$, e.g., randomly}
		\For{$n = 1$ to $N_{\text{max-iter}}$}
		
        \State{Get ${\bm B}^{(n)}_{r} $ using (\ref{bc}) for given  ${\bm C}^{(n-1)}_{r} $}	
        
        \State{Get ${\bm C}^{(n)}_{r} $ using (\ref{cc}) for given  ${\bm B}^{(n)}_{r} $}	
		\EndFor
		\State{Set $\bm{\hat{B}}_{r} = {\bm B}^{(N_{\text{max-iter}})}_{r}$ and $\bm{\hat{C}}_{r} = {\bm C}^{(N_{\text{max-iter}})}_{r}$ }
		\State{Get $\bm{\mathcal{\hat Y}}_{r} = \bm{\mathcal{S}} \times_1  {\bm A}  \times_2  \bm{\hat{B}}^{\mathsf{T}}_{r}  \times_3  \bm{\hat{C}}^{\mathsf{T}}_{r}$, go back to Step (\ref{Step3})}
		\EndFor
		\State{\textbf{Output}: $\bm{\hat{H}}_{\text{SALSA}} = \sum_{r = 1}^{R} \bm{\hat{C}}_{r} \otimes \bm{\hat{B}}_{r} \in\mathbb{C}^{I  \times  J }$ }
	\end{algorithmic}
\end{algorithm}

 Note that, due to the right filtering, the SALSA method in Algorithm~\ref{SALSA} requires that (C1) $I_1 \leq LJ_2$ and (C2) $ I_2 \leq LJ_1$, i.e., $T_\text{BS} \geq \min \Big\{\frac{I_1}{{N}_\text{RF}  J_2} , \frac{I_2}{{N}_\text{RF}  J_1} \Big\}$ to provide an accurate channel estimation. Therefore, under practical settings, the SALSA method in Algorithm~\ref{SALSA} requires less training overhead than the LS method in (\ref{LS}). On the other hand,  assuming that the complexity of calculating the Moore-Penrose pseudo-inverse of an $n\times m$ matrix is on the order of $\mathcal{O}(\min\{n,m\}^3)$, then the complexity of the LS method in (\ref{LS}) is on the order of $\mathcal{O}( \min\{ L ,  J \}  )^3$, while for the SALSA method in Algorithm~\ref{SALSA} the complexity is on the order of $\mathcal{O}(R \cdot N_{\text{max-iter}} \cdot I_1^3 \cdot I_2^3)$, assuming that the  (C1) and (C2) conditions are satisfied.

\section{Simulation Results}\label{SecNumResults}
We adopt the 3GPP clustered delay line (CDL) channel model described in TR 38.901 \cite{3gpp38901}, where a step-by-step tutorial of it along the MATLAB scripts for channel generation is presented in \cite{CDLChannel}. Specifically, in our simulation, we first generate a time-domain channel tensor $\bm{\mathcal{H}} \in \mathbb{C}^{N_\text{BS} \times N_\text{UE} \times N_\text{taps}}$, where $N_\text{taps}$ represents the number of time-domain channel taps calculated according to \cite[Eqn. (64)]{CDLChannel} and using the system parameters shown in Table~\ref{TableSys}. Then, we perform a $N_\text{SC}$-point FFT operation along the third dimension for each receive-transmit antenna pair to obtain the frequency-domain channel tensor $\bm{\mathcal{H}} \in \mathbb{C}^{N_\text{BS} \times N_\text{UE} \times N_\text{SC}}$, where the $k$th slice matrix $\bm{H}_k = \bm{\mathcal{H}}_{[:,:,k]} \in \mathbb{C}^{N_\text{BS} \times N_\text{UE}}$ represents the the $k$th subcarrier frequency-domain MIMO channel matrix. 

We show the simulation results in terms of the normalized mean-square-error (NMSE) that is defined as $\text{NMSE} = \mathbb{E}\{ \|{\bm H} - \bm{\hat{H}}_\text{X}\|_{\text{F}}^2\}/\mathbb{E} \|{\bm H} \|_{\text{F}}^2\}$, where $\text{X} \in \{\text{LS, SALSA} \}$. The signal-to-noise ratio (SNR) is defined as $\text{SNR} = \mathbb{E} \{\| \bm{\mathcal{Y}} - \bm{\mathcal{Z}}\|_{\text{F}}^2\}/\mathbb{E} \{\|\bm{\mathcal{Z}}\|_{\text{F}}^2\}$. In all simulation scenarios, we set $N_\text{BS} = 64$, $N_\text{UE} = 4$, $N_\text{SC} = 16$, $T_\text{UE} = N_\text{UE}$, $N_\text{RF} = 4$, $N_\text{G} = 2$, and assume a random generation of the analog decoding matrix $\bm{A} \in \mathbb{C}^{ T_\text{BS} {N}_\text{RF}  \times N_\text{BS}}$, where every nonzero entry is obtained as $a = {1}/{\sqrt{\mathring{N}_\text{BS}}} \cdot e^{j \phi}$, where $\phi \in [0, 2\pi]$. 

\begin{table}
	\centering 
	\caption{System Parameters}
	\label{TableSys}
	\begin{tabular}{c|c}
		Parameter & Value \\\hline
		Scenario & UMi \\
		Cell radius  & 100 m \\
		BS (UE) height  & 10 (1.5) m \\
		Carrier frequency $f_c$ & 4 GHz \\ 
		Sampling frequency $f_s$ & 30.72 MSamples/s \\
%		Subcarrier spacing & 30 kHZ \\
		No. of subcarriers $N_\text{SC}$ & $16$ \\
	    %$N_{\text{FFT}}$ & $N_\text{SC}$ \\
	    No. of antennas at BS $N_\text{BS}$ &  64 $(8\times 8)$\\
	    No. of antennas at UE $N_\text{UE}$ & 4 $(2\times 2)$\\
	    Polarization & Single \\ \hline
	\end{tabular}
%\vspace{-10pt}
\end{table}

\begin{figure*}
	\centering
	\includegraphics[width=0.47\linewidth]{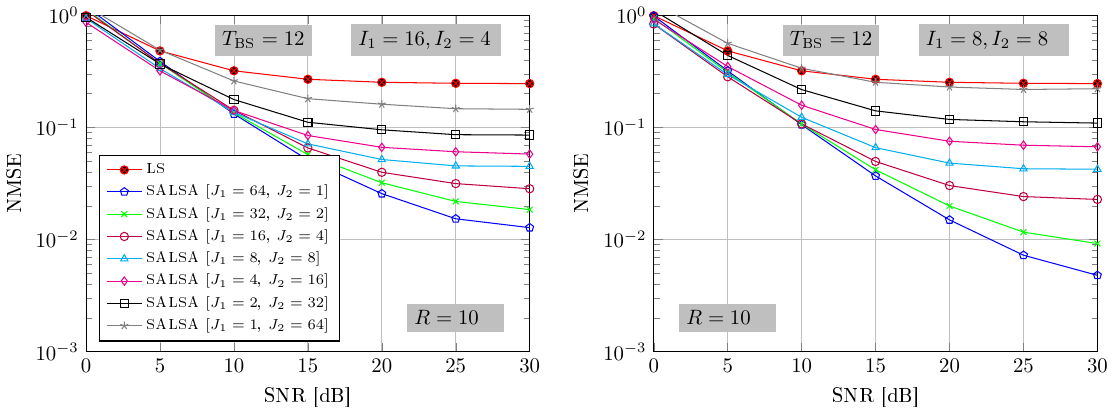}~
	\includegraphics[width=0.47\linewidth]{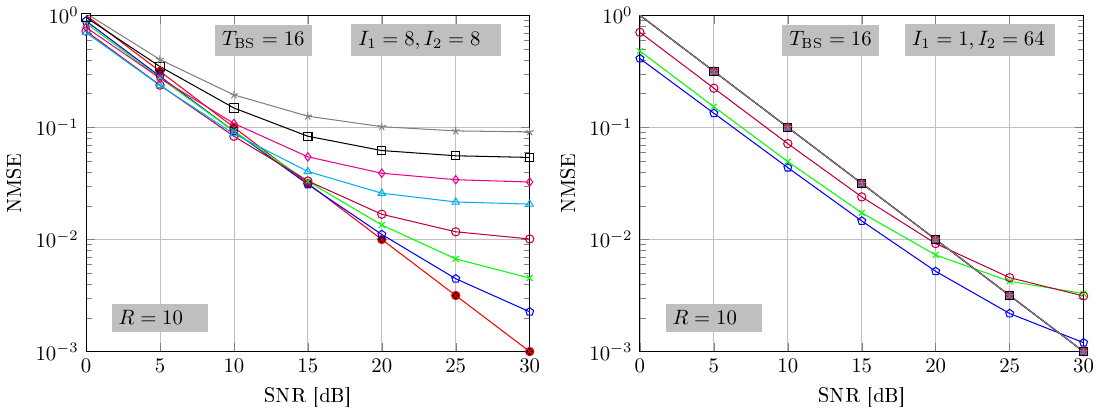}
	\caption{NMSE vs. SNR for different $I$ and $J$ division scenarios.}
	\label{fig1}
\end{figure*}

Initially, we show simulation results investigating the best division scenario of $I$ and $J$ with the constraints of $I = I_1I_2$, $J = J_1J_2$, $I_d\geq 1$, $J_d\geq 1$, and $I_d, J_d$ are Natural numbers, where $d\in \{1,2\}$. Recall that $I = {N}_\text{BS}$ and $J = {N}_\text{UE}N_\text{SC}$. Therefore, we have $I = J = 64$ and the candidate numbers of $I_d$ and $ J_d$ are $1, 2, 4, 8, 16, 32 $, and $64$. Therefore, we have in total 49 different division scenarios as illustrated in Table \ref{Scenarios}. We have simulated the SALSA algorithm using all the 49 possible scenarios. In Fig. \ref{fig1}, we show the NMSE versus SNR results for some selected $I$ and $J$ division scenarios. The other scenarios are not shown, due to space limitations, but we note that their NMSE performance are inferior compared to the shown scenarios. 

From Fig. \ref{fig1}, when $T_\text{BS} = 12$, i.e., $L = T_\text{BS} {N}_\text{RF} = 48 < N_\text{BS}$, the analog training matrix $\bm{A} \in \mathbb{C}^{ L  \times N_\text{BS}}$, i.e., the 1st factor matrix of the measurement tensor in (\ref{Yrten}), is left non-invertible, i.e., $[{\bm A}]^{+} {\bm A} \neq \bm{I}$. Therefore, the LS-based method has a very bad channel estimation accuracy NMSE. On the other hand, we can see that the best NMSE of SALSA method is achieved when $I_1 = 8$, $I_2 = 8$, $J_1 = 64$, and $J_2 = 1$, i.e., when ${\bm B}_r \in \mathbb{C}^{8 \times 64}$ and ${\bm C}_r \in \mathbb{C}^{8 \times 1}, \forall r$. The main reason is that by dividing $I = 64$ equally between $I_1$ and $I_2$, i.e., $I_1 = I_2 = 8$, SALSA reduces the impact of the non-invertibility of $\bm{A}$ by distributing it between the 2nd (i.e., ${\bm B}_r$) and the 3rd (i.e., ${\bm C}_r$) factor matrices of the measurement tensor, which leads to a better channel estimation accuracy. On the other hand, by setting $J_1 = 64$ and $J_2 = 1$, the required number of channel factor matrices $R$ reduces as compared to the other division scenario, as we have illustrated above in Fig.~\ref{fig2}.

Differently, when $T_\text{BS} = 16$, i.e., $L = N_\text{BS}$, the analog training matrix $\bm{A}$ is left invertible, i.e., $[{\bm A}]^{+} {\bm A} = \bm{I}$. Therefore, the LS-based method has an accurate channel estimation accuracy. For SALSA method, on the other hand, we can see that when $I_1 = 8$ and $I_2 = 8$, the estimation accuracy of SALSA improves as we increase $J_1$ and decrease $J_2$, where the best result is obtained when we have $J_1 = 64$ and $J_2 = 1$, i.e., similar to the case above when $T_\text{BS} = 12$. Nonetheless, we can see that the SALSA method can obtain a more accurate channel estimation, compered to the LS-based method, by setting $I_1 = 1 $, $ I_2 = 64$, $J_1 = 64$, and $J_2 = 1$, i.e., ${\bm B}_r \in \mathbb{C}^{1 \times 64}$ and ${\bm C}_r \in \mathbb{C}^{64 \times 1}$ (or, \textit{not shown in the figure},  by setting $I_1 = 64 $, $ I_2 = 1$, $J_1 = 1$, and $J_2 = 64$, i.e., ${\bm B}_r \in \mathbb{C}^{64 \times 1}$ and ${\bm C}_r \in \mathbb{C}^{1 \times 64}$). In the both these scenarios, the channel matrix $\bm{H} \in \mathbb{C}^{64\times 64}$ in (\ref{HCB}) is decomposed into a summation of $R$ factor matrices ${\bm B}_r \otimes {\bm C}_r \in \mathbb{C}^{64\times 64}$, each having a rank-one, i.e., \textsf{rank}$\{{\bm B}_r \otimes {\bm C}_r\}= 1, \forall r$, which leads to a better channel estimation accuracy.

% \usepackage{multirow}
\begin{table}[]
		\centering 
	\caption{Division scenarios of $I_1, I_2, J_1$, and $J_2$ }
	\label{Scenarios}
	\begin{tabular}{|c|c|c|}
		\hline
			Scenario No.   &  $I_1 $ and $ I_2$    values              & $J_1 $ and $ J_2$    values         \\ \hline
			Scenario 1  & \multirow{3}{*}{$[I_1, I_2] = [64, 1]$} & $[J_1, J_2] = [64, 1]$        \\ \cline{1-1} \cline{3-3} 
		&                    & $\vdots$ \\ \cline{1-1} \cline{3-3} 
		Scenario 7  &                    &       $[J_1, J_2] = [1, 64]$   \\ \hline
	    $\vdots$	& $\vdots$ & $\vdots$ \\ \hline
		Scenario 43 & \multirow{3}{*}{$[I_1, I_2] = [1, 64]$} &    $[J_1, J_2] = [64, 1]$      \\ \cline{1-1} \cline{3-3} 
		&                    & $\vdots$ \\ \cline{1-1} \cline{3-3} 
		Scenario 49 &                    &     $[J_1, J_2] = [1, 64]$     \\ \hline
	\end{tabular}
\end{table}

\begin{figure}
	\centering
	\includegraphics[width=0.75\linewidth]{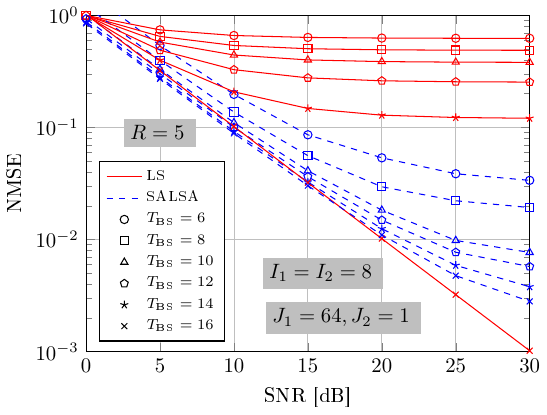} \\
	\caption{NMSE vs. SNR with varying $T_{\text{BS}}$.}
	\label{fig3}
\end{figure}

\begin{figure}
	\centering
	\includegraphics[width=0.75\linewidth]{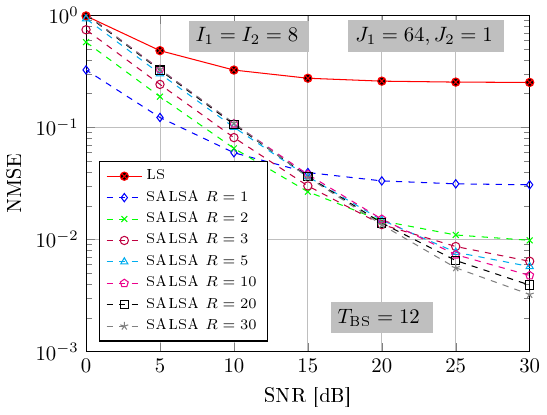} \\
	\caption{NMSE vs. SNR with varying $R$.}
	\label{fig4}
\end{figure}

In Figs. \ref{fig3} and \ref{fig4} we show NMSE versus SNR simulation results with varying the number of channel training overhead, i.e., $T_{\text{BS}}$ and the number of channel factor matrices, i.e., $R$, respectively. From Fig. \ref{fig3}, we can see that the channel estimation accuracy of both methods, i.e., LS-based and SALSA improves as $T_{\text{BS}}$ increases. However, SALSA significantly outperforms LS-based with all $T_{\text{BS}} < 16$, i.e., $L< N_\text{BS}$ scenarios, wherein the analog training matrix $\bm{A} \in \mathbb{C}^{ L  \times N_\text{BS}}$ is left non-invertible. 

On the other hand, we can see from Fig. \ref{fig4} that the SALSA channel estimation accuracy increases with the increasing $R$, in the high SNR regime, while it decreases with the increasing $R$, in the low SNR regime. The main reason is that, in the high SNR regime, the noise impact is minimal and by increasing $R$, the channel estimation accuracy increases, as we have illustrated above in Fig.~\ref{fig2}. On the other hand, in the low SNR regime, the channel measurement tensor is noise-limited and, therefore, the impact of noise increases by increasing $R$, i.e., after a certain $R$, the estimated channel factor matrices are very noisy that decreases the overall estimation accuracy. Clearly, for every SNR regime/level, there is an optimal $R$ value, wherein the channel estimation accuracy is maximized, which we leave for a follow up future work.      

\vspace{-10pt}
\section{Conclusion}
In this paper, we have proposed a novel channel estimation method for MIMO-OFDM sub-6 GHz communication systems called SALSA. We have shown that an accurate channel estimation can be obtained with a small training overhead by exploiting a hidden tensor structure in the received measurement matrix, which estimates the channel matrix sequentially using an ALS-based method. Our results show that the SALSA method outperforms the conventional LS-based method, especially in the low training overhead, which makes it more appealing for practical implementations.

%\pagebreak
%\newpage
\bibliographystyle{IEEEtran}
\bibliography{references}
\end{document}